\documentclass[aps,prl,twoside,twocolumn,floatfix,amsmath,showpacs,amssymb]{revtex4}

 \usepackage[latin1]{inputenc}
 \usepackage{amsmath}
 \usepackage{subfigure}
 \usepackage{placeins}
 \usepackage{floatflt}
 \usepackage{amsmath}
 \usepackage{amssymb}
 \usepackage{wrapfig}
 \usepackage[]{tocbibind}
 \usepackage[all]{xy}

 \usepackage{ifthen}
 \usepackage{times}
 
\makeatletter
\newcommand\figcaption{\def\@captype{figure}\caption}
\newcommand\tabcaption{\def\@captype{table}\caption}
\makeatother

\usepackage[pdftex]{graphicx}                                

\begin{document}

\DeclareGraphicsExtensions{.jpg,.pdf,.png,.mps,.eps,.ps}             

\title{Deep sub-threshold $\mathbf{\Xi^-}$ production in Ar+KCl reactions at 1.76\boldmath{$A$}~GeV}

\author{
G.~Agakishiev$^{8}$, A.~Balanda$^{3}$, R.~Bassini$^{9}$, 
D.~Belver$^{15}$, A.V.~Belyaev$^{6}$, A.~Blanco$^{2}$, M.~B\"{o}hmer$^{11}$, J.\,L.~Boyard$^{13}$, 
P.~Braun-Munzinger$^{4}$, P.~Cabanelas$^{15}$
% ...
\\
(HADES collaboration)
}

\affiliation{
\\\mbox{$^{1}$Istituto Nazionale di Fisica Nucleare - Laboratori Nazionali del Sud, 95125~Catania, Italy}\\
\mbox{$^{2}$LIP-Laborat\'{o}rio de Instrumenta\c{c}\~{a}o e F\'{\i}sica Experimental de Part\'{\i}culas , 3004-516~Coimbra, Portugal}\\
\mbox{$^{3}$Smoluchowski Institute of Physics, Jagiellonian University of Cracow, 30-059~Krak\'{o}w, Poland}\\
\mbox{$^{4}$GSI Helmholtzzentrum f\"{u}r Schwerionenforschung GmbH, 64291~Darmstadt, Germany}\\
\mbox{$^{5}$Institut f\"{u}r Strahlenphysik, Forschungszentrum Dresden-Rossendorf, 01314~Dresden, Germany}\\
\mbox{$^{6}$Joint Institute of Nuclear Research, 141980~Dubna, Russia}\\
\mbox{$^{7}$Institut f\"{u}r Kernphysik, Johann Wolfgang Goethe-Universit\"{a}t, 60438 ~Frankfurt, Germany}\\
\mbox{$^{8}$II.Physikalisches Institut, Justus Liebig Universit\"{a}t Giessen, 35392~Giessen, Germany}\\
\mbox{$^{9}$Istituto Nazionale di Fisica Nucleare, Sezione di Milano, 20133~Milano, Italy}\\
\mbox{$^{10}$Institute for Nuclear Research, Russian Academy of Science, 117312~Moscow, Russia}\\
\mbox{$^{11}$Physik Department E12, Technische Universit\"{a}t M\"{u}nchen, 85748~M\"{u}nchen, Germany}\\
\mbox{$^{12}$Department of Physics, University of Cyprus, 1678~Nicosia, Cyprus}\\
\mbox{$^{13}$Institut de Physique Nucl\'{e}aire (UMR 8608), CNRS/IN2P3 - Universit\'{e} Paris Sud, F-91406~Orsay Cedex, France}\\
\mbox{$^{14}$Nuclear Physics Institute, Academy of Sciences of Czech Republic, 25068~Rez, Czech Republic}\\
\mbox{$^{15}$Departamento de F\'{\i}sica de Part\'{\i}culas, Univ. de Santiago de Compostela, 15706~Santiago de Compostela, Spain}\\
\mbox{$^{16}$Instituto de F\'{\i}sica Corpuscular, Universidad de Valencia-CSIC, 46971~Valencia, Spain}\\
\mbox{$^{a}$ also at ISEC Coimbra, ~Coimbra, Portugal}\\
\mbox{$^{b}$ also at Technische Universit\"{a}t Dresden, 01062~Dresden, Germany}\\
\mbox{$^{c}$ also at Dipartimento di Fisica, Universit\`{a} di Milano, 20133~Milano, Italy}\\
\mbox{$^{d}$ also at Excellence Cluster Universe, Technische Universit\"{a}t M\"{u}nchen, 85748 Garching, Germany}\\
}
\date{\today}

\begin{abstract}
We report first results on a deep sub-threshold production of the doubly strange hyperon  
$\Xi^-$ in a heavy-ion reaction. At a beam energy of 1.76$A$~GeV the reaction Ar+KCl 
was studied with the High Acceptance Di-Electron Spectrometer 
(HADES) at SIS18/GSI. A high-statistics and high-purity 
$\Lambda$ sample was collected, allowing for the investigation of the decay channel  
$\Xi^- \rightarrow \Lambda \pi^-$. 
The deduced $\Xi^-/(\Lambda+\Sigma^0)$ production ratio of 
$(5.6 \pm 1.2 \, ^{+1.8}_{-1.7})\cdot 10^{-3}$
is significantly larger than available model predictions. 
\end{abstract}

\pacs{25.75.-q, 25.75.Dw} 

\maketitle 
The doubly strange $\Xi^-$ baryon (also known as cascade particle) 
has, in vacuum, a mass of 1321.3\,MeV and decays ($\mathrm{c} \tau = 4.91$\,cm) 
almost exclusively into the $\Lambda$-$\pi^-$ final state \cite{PDG}. 
In elementary nucleon-nucleon (NN) collisions near threshold 
it must be co-produced with two kaons 
ensuring strangeness conservation. This requires a minimum beam energy of 
$E_{thr}=3.74$~GeV ($\sqrt{s_{thr}}=3.25$~GeV). 
In heavy-ion collisions, the $\Xi^-$ yield was measured at various beam energies covered by the 
RHIC \cite{RHIC07}, SPS \cite{SPS04_NA57,SPS08_NA49} and AGS \cite{AGS04} accelerators.  
Though cooperative processes in heavy-ion reactions allow for particle production below NN 
threshold, no sub-threshold $\Xi^-$ production was observed so far. 
Predictions of sub-threshold cascade production  
at energies available with the heavy-ion synchrotron SIS18 at GSI, Darmstadt,  
were presented within a relativistic transport model 
\cite{CheMingKo04}. The cross sections of the strangeness exchange reactions 
$\mathrm{\bar{K} Y \rightarrow \pi \Xi }$ ($\mathrm{Y=\Lambda,\,\Sigma}$), 
which are essential for $\Xi$ creation below the nucleon-nucleon threshold, were taken from  
a coupled-channel approach based on a flavor SU(3)-invariant hadronic Lagrangian 
\cite{CheMingKo02}.
The $\Xi^-/\Lambda$ ratio was found to amount to a few times $10^{-4}$, 
varying with system size and beam energy, however, being fairly independent on centrality. 
At SIS18 energies, also $\mathrm{\bar{K}}$ production, 
being a prerequisite of the above strangeness exchange reactions,  
proceeds below the NN threshold ($E_{thr}=2.5$~GeV). A similar strangeness exchange reaction 
like that relevant for $\Xi$ production 
is found to be dominant in sub-threshold $\mathrm{\bar{K}}$ 
production in heavy-ion collisions  
\cite{Hartnack03}, i.e.\ $\mathrm{\pi Y \rightarrow \bar{K} B}$ ($\mathrm{B=N,\,\Delta}$). 
Thus, medium effects on strange meson properties, like effective anti-kaon masses and hence 
reduced production thresholds, could strongly influence the $\Xi$ yield. 
However, the authors of \cite{CheMingKo04} found the $\Xi$ yield to be more 
sensitive to the magnitude of the cross sections of strangeness-exchange reactions than 
to the medium effects due to modified kaon properties. 
Generally, the yield of multi-strange particles, measured below their production 
threshold in NN collisions, is expected to be sensitive to the equation of state (EoS) 
of nuclear matter. In heavy-ion reactions, the necessary energy for the 
production of these particles is accumulated via multiple collisions involving nucleons, 
produced particles and short-living resonances. The corresponding number of such collisions 
increases with the density inside the reaction zone the maximum of which in turn depends  
on the stiffness of the EoS. 

In this Letter we report on the first observation of sub-threshold $\Xi^-$ production 
in heavy-ion collisions. The experiment was performed with the 
{\bf H}igh {\bf A}cceptance {\bf D}i-{\bf E}lectron {\bf S}pectrometer (HADES) 
at SIS18 \cite{hadesSpectro}.
HADES, primarily designed to measure di-electrons \cite{HADES-PRL07}, 
offers excellent hadron identification capabilities, 
too \cite{PhD_Schmah,Fabbietti_SQM08,hades_kpm_phi_09}. 

A $^{40}$Ar beam of about $10^6$ particles/s with kinetic energy of 1.756$A$~GeV 
($\sqrt{s_{NN}}=2.61$~GeV) was incident on a four-fold segmented $^{\mathrm{nat}}$KCl 
target with a total thickness of 5\,mm corresponding to $3.3\,\%$ interaction probability. 
The beam energy is known with a precision of about $10^{-3}$. 
The energy loss of the beam particles in the target is estimated to be 
less than 0.5\,$A$MeV per target slice. 
The position resolution of the primary (reaction) vertex amounts to 0.3\,mm in both 
transverse directions while in beam direction it amounts to 1.5\,mm as expected from 
the finite thickness of the target slices. 
The data readout was started by a first-level trigger (LVL1) decision, 
requiring a minimum charged-particle multiplicity $\ge 16$
in the time-of-flight detectors. 
The integrated cross section selected by this trigger 
comprises approximately the most central 35\,\% of the total reaction cross section.  
About 700 million LVL1 events were processed for the present $\Xi^-$ investigation.

In the present analysis we identified the $\Lambda$ hyperons 
through their decay $\Lambda \rightarrow \mathrm{p} \pi^-$. 
Note that the reconstructed $\Lambda$ yield includes the decay $\Lambda$'s of the  
(slightly heavier) $\Sigma^0$ baryon decaying exclusively into $\Lambda$ and a 
photon \cite{PDG} which cannot be detected with HADES. Hence, the 
$\Lambda$ yield has to be understood as that of $(\Lambda+\Sigma^0)$ throughout the paper. 
To allow for $\Lambda$ selection various cuts on single-particle and two-particle 
quantities were applied. The most important ones act on geometrical distances, 
i.e.\ i) minimum values of the p, $\pi^-$ track distances to the primary vertex 
(\mbox{p-VecToPrimVer}, \mbox{$\pi_1$-VecToPrimVer}), 
ii) an upper threshold of the p-$\pi^-$ minimum track distance 
(\mbox{p-$\pi_1$-MinVecDist}), and 
iii) a minimum value of the $\Lambda$ vertex distance to the primary vertex 
(\mbox{$\Lambda$-VerToPrimVer}). 

\begin{figure}[!htb]
\begin{center}
\includegraphics[width=0.9\linewidth,viewport=0 0 570 260]{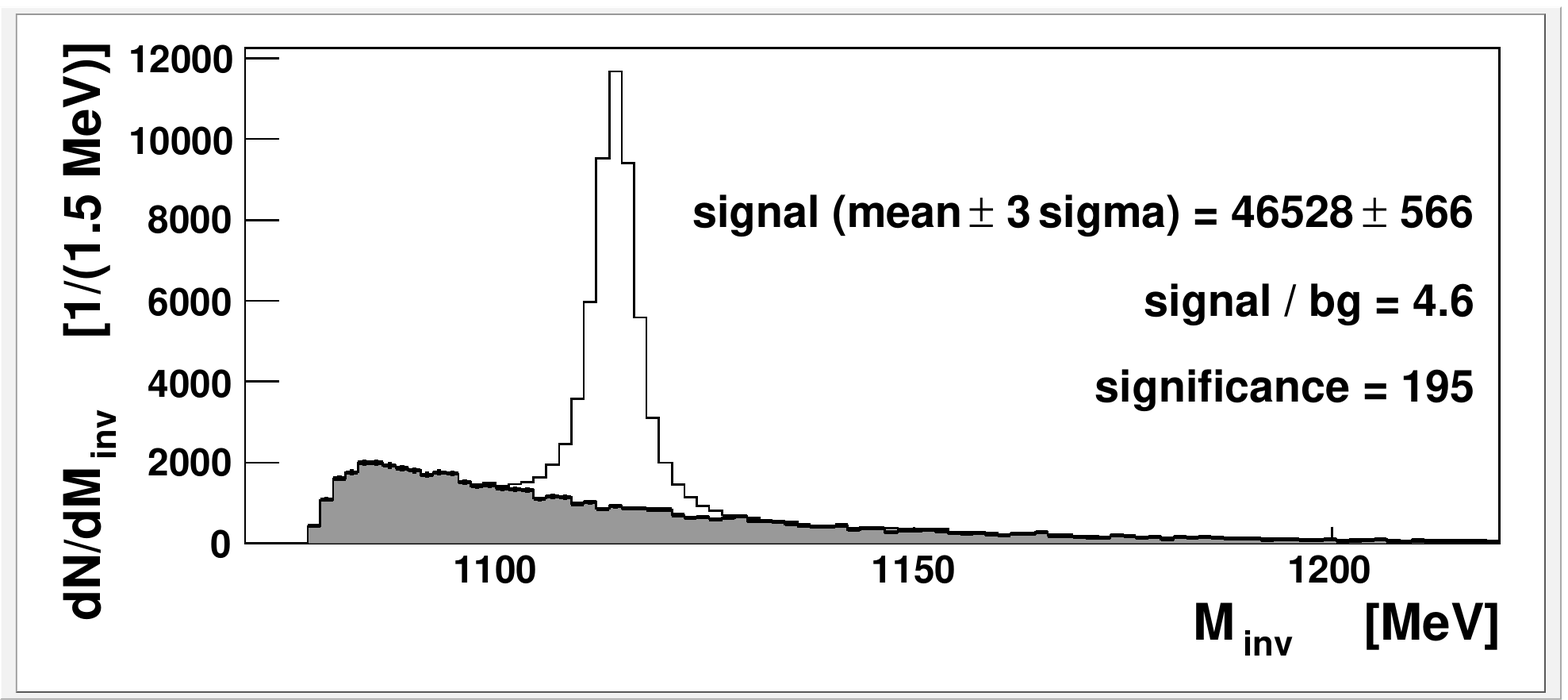}
\caption[]{The p-$\pi^-$ invariant mass distribution. 
Hatched histogram: Scaled combinatorial background produced via event mixing. 
\label{lambda_inv_mass}}
\end{center}
\end{figure}
With these conditions we first analysed the invariant-mass distribution of 
proton-$\pi^-$ pairs (Fig.\,\ref{lambda_inv_mass}). A clear $\Lambda$ signal could be separated 
from the combinatorial background (bg) as determined via the event mixing technique. 
The background normalization was performed over the given invariant-mass range, 
except a $\pm15$\,MeV interval around the $\Lambda$ peak. 
For a $\pm 3 \sigma$ mass cut around the $\Lambda$ peak,  
the signal-to-background ratio and the significance, 
defined as $\mathrm{signal/\sqrt{signal+bg}}$, amount to 4.6 and 195, respectively. 
The total $\Lambda$ yield for the given cuts is $N_{\Lambda}=46,500\pm 600$. 
Fitting a Gaussian function to the signal, we obtain a mean value of 
$(1114.2 \pm 0.1)$\,MeV and a width ($\sigma$) of $(2.7 \pm 0.1)$\,MeV.  

Taking this high-statistics $\Lambda$ sample, we
started the $\Xi^-$ investigation by combining - for each event containing a 
$\Lambda$ candidate - the $\Lambda$ with those $\pi^-$ mesons not already 
contributing to the $\Lambda$. The result was  
a structureless $\Lambda$-$\pi^-$ invariant mass distribution.   
Hence, additional conditions were necessary: 
iv) a lower limit on the 2nd $\pi^-$ (potential $\Xi^-$ daughter) 
track distance to primary vertex (\mbox{$\pi_2$-VecToPrimVer}), 
v) an upper limit of the distance of the $\Xi^-$ pointing vector w.r.t. the primary vertex 
(\mbox{$\Xi$-VecToPrimVer}), 
vi) a maximum value of the minimum track distance of the $\Lambda$ and the 2nd $\pi^-$ 
(\mbox{$\pi_2$-$\Lambda$-MinVecDist}),  
vii) a minimum value of the distance of the $\Xi^-$ vertex relative to the primary one 
(\mbox{$\Xi$-VerToPrimVer}), and 
viii) a window of $\pm 7$\,MeV around the $\Lambda$ mass peak of the p-$\pi^-$ invariant 
mass distribution. 

\begin{figure}[!htb]
\begin{center}
\includegraphics[width=0.9\linewidth,viewport=0 0 570 520]{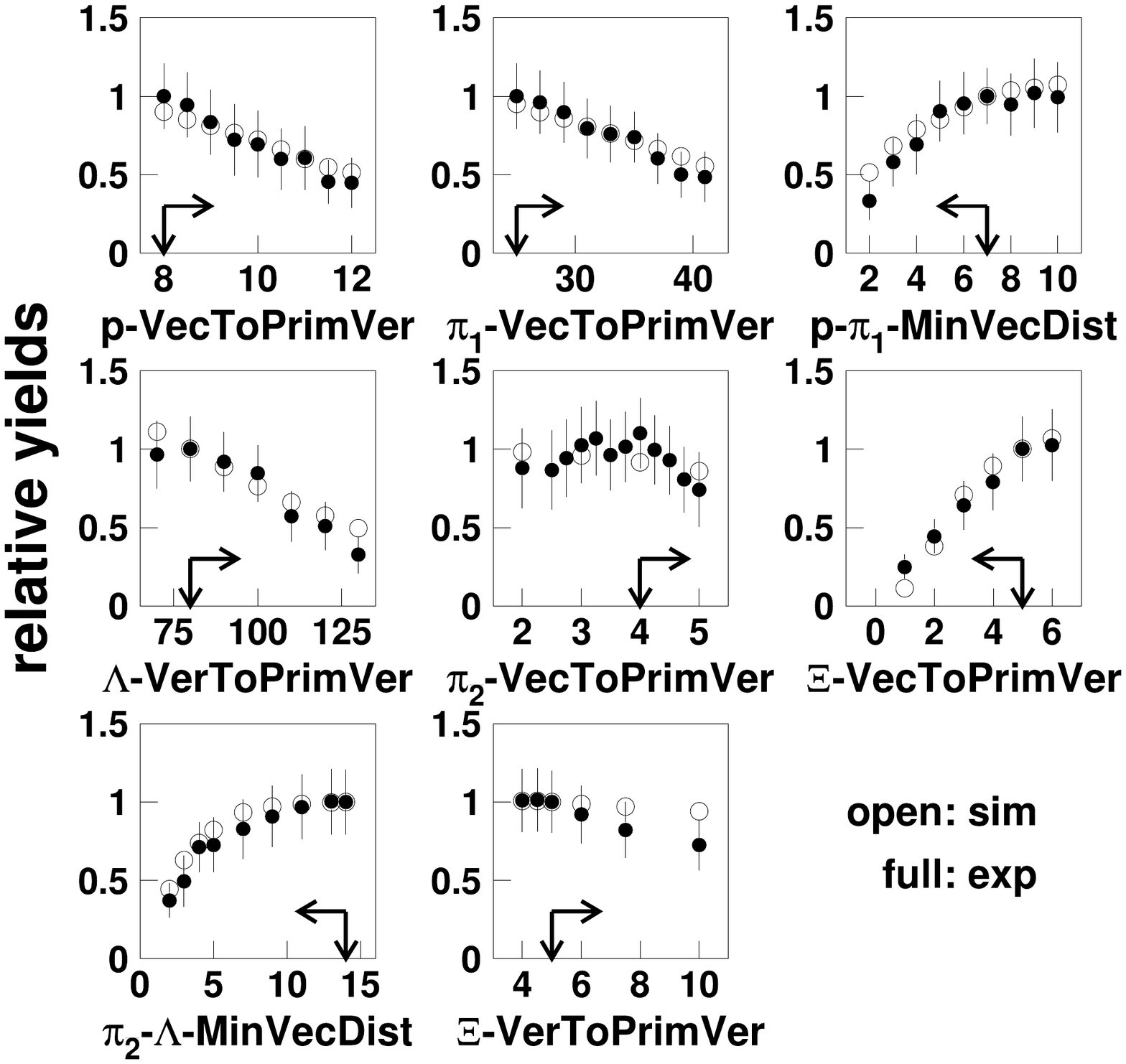}
\caption[]{Relative $\Xi^-$ yields as a function of the cut value of  
various $\Lambda$ and $\Xi^-$ geometrical 
distances (see text, units are mm). The full (open) dots display the experimental  
(simulation) data. The vertical and horizontal arrows indicate the chosen cut values 
and the region of accepted distances, respectively. \vspace*{-5mm}  
\label{xi_eff_vs_cuts_exp_sim} }
\end{center}
\end{figure}
The conditions on the geometrical quantities are summarized in 
Fig.\,\ref{xi_eff_vs_cuts_exp_sim}, where the optimum cut values are indicated by arrows.
We studied the stability of the signal if more stringent conditions were chosen. 
Because of the limited statistics all other cuts were kept fixed at the optimum values 
when varying a single cut quantity. The dependences on the various geometrical distances of  
experimental data and GEANT \cite{GEANT} simulations (see below) are found in good agreement. 

Figure\,\ref{xi_inv_mass} shows the invariant mass distribution of $\Lambda$-$\pi^-$ pairs 
after applying all conditions. 
Indeed, a narrow signal shows up on top of a smooth distribution. For an invariant-mass window of 
$\pm 10$\,MeV (4 bins) around the peak center, we find $N_{\Xi^-}=141\pm 31 \pm 25$ entries 
to be attributed to $\Xi^-$ with the given statistical and systematic errors. 
The signal-to-background ratio and the significance 
amount to 0.17 and 4.6, respectively.  
The given systematic error of the signal is due to the signal variation for various 
histogram binnings, background normalization regions and mass windows assigned to the signal. 
These systematic variations are also reflected in the significance of the signal of about $4-6$. 
The full line in the bottom panel of Fig.\,\ref{xi_inv_mass} 
represents a Gaussian fit to the signal. The mean value of 
$(1320 \pm 1)$\,MeV is well in agreement with the PDG value of 1321.3\,MeV \cite{PDG}. 
Taking into account the bias due to the rather large bin size of 5\,MeV to be used for 
statistical reasons, the peak width ($\sigma$) of $(4 \pm 1)$\,MeV is in fair 
agreement with GEANT simulations which predict for $\Lambda$ and $\Xi^-$ 
baryons almost equal values of about 2.5\,MeV. 
In order to ensure that not a fake signal is selected, we performed a $\Lambda$-side-band 
analysis.  No signal was found when choosing instead of condition viii) a window in the 
p-$\pi^-$ invariant mass of 
$\mathrm{ 10< \vert M_{p \pi^-}-\langle M_{\Lambda}\rangle \vert < 25}$\,MeV. 
Furthermore, when dividing randomly the data sample into two sub-samples,  
the $\Xi^-$ sub-yields were found - within errors - compatible with 
half of the above quoted total yield. 
\begin{figure}[!htb]
\begin{center}
\includegraphics[width=0.9\linewidth,viewport=0 0 570 540]{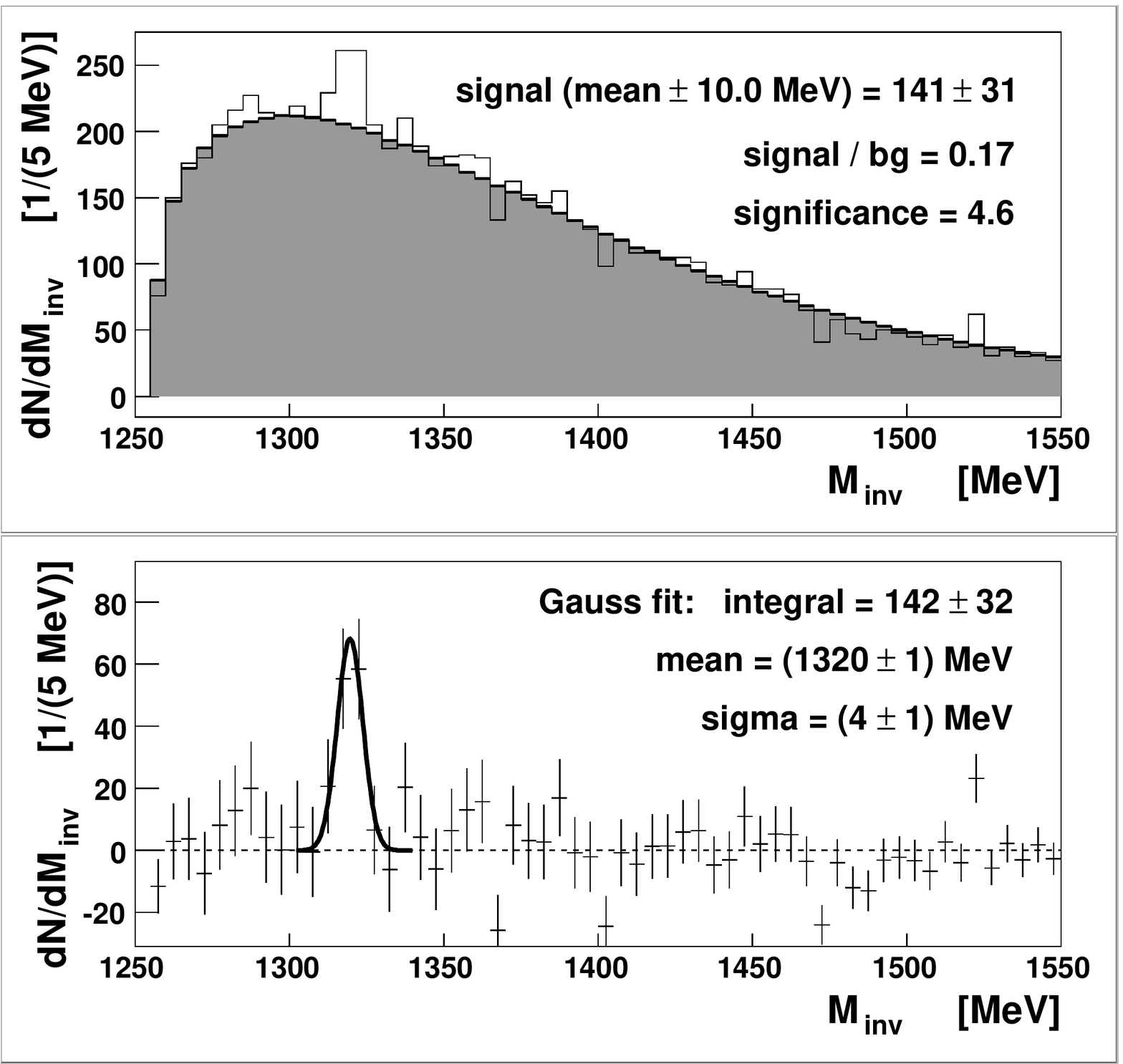}
\caption[]{Top: The same as Fig.\,\ref{lambda_inv_mass}, but for $\Lambda$-$\pi^-$ pairs. 
Bottom: The invariant mass distribution after background subtraction. The full line represents 
a Gaussian fit to the $\Xi^-$ signal.\vspace*{-5mm}  
\label{xi_inv_mass} }
\end{center}
\end{figure}

Corrections for the finite acceptances and reconstruction efficiencies were deduced from 
simulations. Thermal $\Lambda$'s ($\Xi^-$'s), characterized by 
the temperature parameter $T_{\Lambda}$ ($T_{\Xi^-}$)  
were generated with the event generator Pluto \cite{Pluto}. 
The experimental $\Lambda$ rapidity distribution 
is found slightly broader than the thermal model distribution 
\cite{PhD_Schmah,Merschmeyer_07}. 
Consequently, in Pluto we allowed also for anisotropic, 
i.e.\ longitudinally elongated, phase-space distributions. For this purpose, 
an additional width parameter of Gaussian rapidity distributions,  
$\sigma_y$, is taken into account. 
The $\Lambda$ parameters are chosen such that the simulation   
reproduces both, the experimental values of the effective inverse slope parameter at mid-rapidity, 
$T_{eff,\,\Lambda}=95$\,MeV, and the rapidity width,  
$\sigma_{y,\,\Lambda}=0.42$ \cite{PhD_Schmah,Fabbietti_SQM08}. 
Since the phase-space distribution of the $\Xi^-$ is not known,  
we investigated its geometrical acceptance for a broad range of the transverse 
and longitudinal shape parameters, i.e.\ the inverse slope 
and rapidity width, $T_{eff,\,\Xi^-}=(95\pm25)$\,MeV and $\sigma_{y,\,\Xi^-}=0.34\pm0.09$, 
respectively. Here, the $\Lambda$ inverse slope serves as a reference value. 
The lower limit of $T_{eff,\,\Xi^-}$ matches the measured inverse slope parameter of 
$\mathrm{K^-}$ mesons \cite{Fabbietti_SQM08,hades_kpm_phi_09} being essential for 
producing $\Xi$ hyperons via strangeness-exchange processes (see above),  
while the upper limit is set by a similar interval above the $\Lambda$ slope. 
We assumed the rapidity width of $\Xi^-$'s to be 
larger than the width of thermal $\Xi^-$'s with a temperature of 95\,MeV but  
smaller than that of $\Lambda$'s.  This choice is substantiated by two facts: Firstly,  
for a thermal rapidity distribution, the width approximately scales with the square root 
of mass ($\sigma_y=\sqrt{T/m_0 c^2}$), and secondly, the cascade particle may carry less 
longitudinal momentum than the $\Lambda$ hyperon, since it contains only 
one light quark which arises from projectile nucleons. 
With the given parameters and their ranges, we calculated the HADES acceptance 
(including the branching ratio for the decay $\Lambda \rightarrow \mathrm{p} \pi^-$) 
for the $\Lambda$ to 
$\epsilon_{\mathrm{acc}, \Lambda}=0.160\pm 0.009$, and for the $\Xi^-$ to 
$\epsilon_{\mathrm{acc}, \Xi^-}=(9.6 ^{+2.3}_{-2.1})\cdot10^{-2}$.  
The simulation data are processed through GEANT modeling the detector response. 
The GEANT data were embedded into 
real experimental data and processed through the full analysis chain. 
Relating the outputs after cuts to the corresponding inputs, 
the $\Lambda$ and $\Xi^-$ reconstruction efficiencies were estimated to  
$\epsilon_{\mathrm{eff}, \Lambda}= (6.1 \pm 0.3) \cdot 10^{-2}$ and 
$\epsilon_{\mathrm{eff}, \Xi^-  }= (5.5 \pm 0.5) \cdot 10^{-2}$, 
respectively. 
We proved our acceptance and efficiency corrections by extracting the lambda yield 
\cite{PhD_Schmah} which is found to be in agreement with existing data \cite{Merschmeyer_07}. 
With the above correction factors, the ratio of $\Xi^-$ and $\Lambda$ 
production yields can be determined. Such a ratio, when derived from the 
same data analysis, has the advantage that systematic errors cancel to a large extent. 
The ratio is calculated as  
\begin{equation}
\hspace*{-2mm}
\frac{P_{\Xi^-}}
     {P_{\Lambda + \Sigma^0}} = 
\frac{N_{\Xi^-}}
     {N_{\Lambda}} \,
\frac{\epsilon_{\mathrm{acc}, \Lambda }}
     {\epsilon_{\mathrm{acc}, \Xi^-   }} \, 
\frac{\epsilon_{\mathrm{eff}, \Lambda }}
     {\epsilon_{\mathrm{eff}, \Xi^-   }} 
=(5.6 \pm 1.2 \, ^{+1.8}_{-1.7}) 10^{-3},
\label{xi_lambda_ratio}
\end{equation}
where statistical and systematic errors are given,   
resulting from adding the individual ones quadratically.
The statistical error in (\ref{xi_lambda_ratio}) is dominated 
by the 20\,\% error of the $\Xi^-$ signal while the systematic error is governed 
by the stability of the signal against cut and background variation and  
by the range of the parameters $T_{\Xi^-}$ and $\sigma_{y,\,\Xi^-}$ entering the 
simulation.  

\begin{figure}[!htb]
\begin{center}
\includegraphics[width=0.75\linewidth,viewport=50 160 570 700]{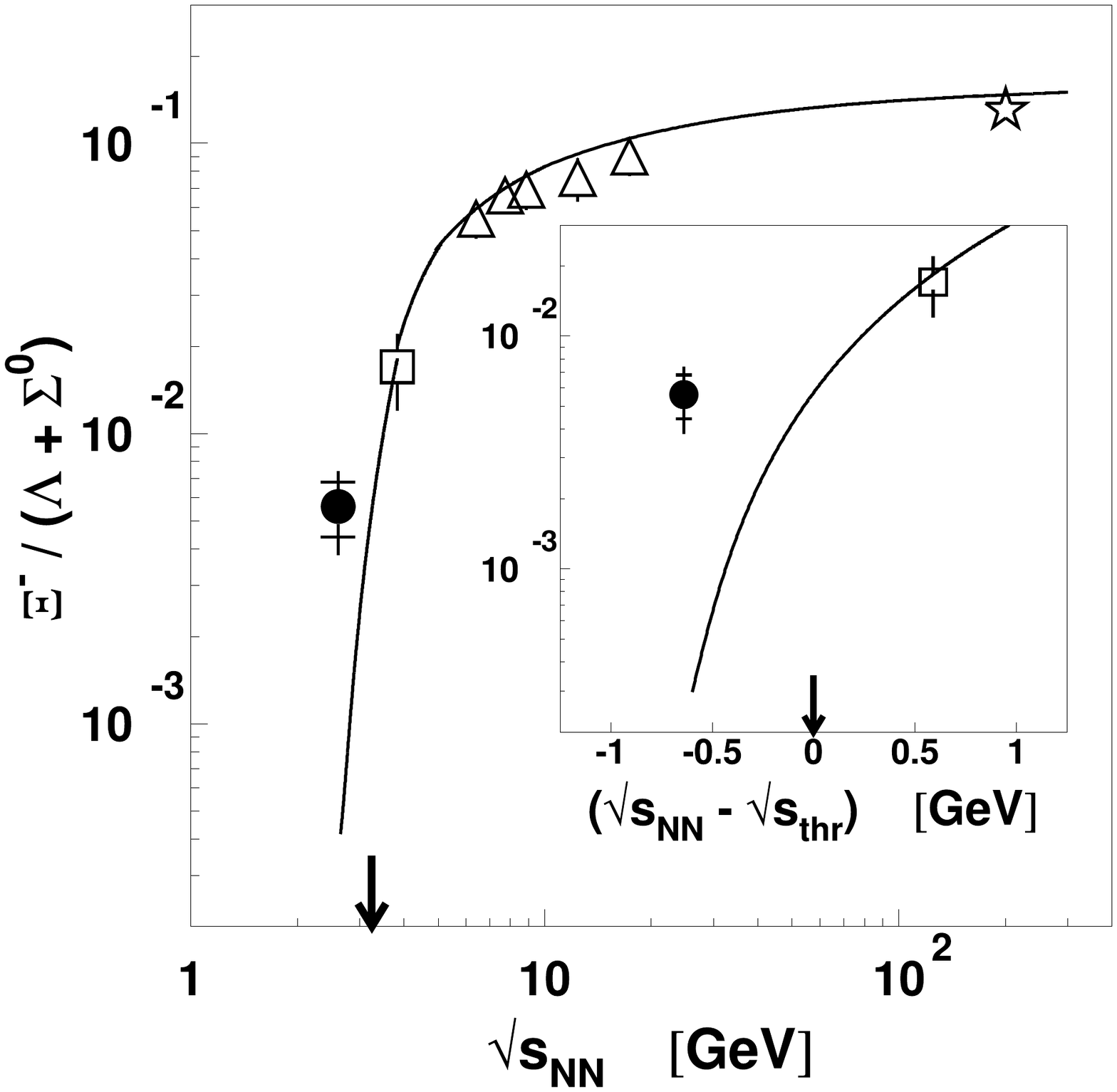}
\caption[]{The yield ratio $\Xi^-/(\Lambda+\Sigma^0)$ as a function of $\sqrt{s_{NN}}$ or 
$\sqrt{s_{NN}} - \sqrt{s_{thr}}$ (inset). The arrow 
gives the threshold in free NN collisions. The open star, triangles and square represent  
data for central Au+Au and Pb+Pb collisions 
measured at RHIC \cite{RHIC07}, SPS \cite{SPS04_NA57,SPS08_NA49} and AGS \cite{AGS04}, 
respectively. The filled circle shows the present ratio (\ref{xi_lambda_ratio}) 
for Ar+KCl reactions at 1.76$A$~GeV (statistical error within ticks, 
systematic error as bar). 
Full line: Statistical model for Au+Au \cite{AndronicPBMRedlich06}. \vspace*{-5mm} 
\label{xi_exc_fct} }
\end{center}
\end{figure}
The deduced $\Xi^-/\Lambda$ ratio (\ref{xi_lambda_ratio})
may be compared with the corresponding ratios at higher 
energies \cite{RHIC07,SPS04_NA57,SPS08_NA49,AGS04}. 
Figure\,\ref{xi_exc_fct} shows a compilation of $\Xi^-/\Lambda$ ratios as a function 
of $\sqrt{s_{NN}}$. The displayed data represent 
the most central 5-10\,\% of collisions of Au+Au or Pb+Pb. At RHIC and SPS energies hardly 
any centrality dependence of the $\Xi^-/\Lambda$ ratio was observed \cite{RHIC07,SPS04_NA57}. 
So far, the lowest energy at which a 
$\Xi^-/\Lambda$ ratio is available is $\sqrt{s_{NN}}=3.84$~GeV, 
i.e.\ an excess energy of +600\,MeV above the NN threshold \cite{AGS04}. The corresponding 
ratio, measured at the AGS at a beam energy of 6$A$~GeV, 
is found to increase slightly with centrality. For central (semi-central) collisions it 
is about three (two) times larger than our value. 
Indeed, a steep decline of the $\Xi^-/\Lambda$ production ratio 
is expected below threshold, where now the first data point is available. 
This allows for comparisons to model calculations. 

The $\Xi^-/\Lambda$ ratio has been estimated within a 
statistical approach \cite{AndronicPBMRedlich06}. 
While RHIC \cite{RHIC07}, SPS \cite{SPS04_NA57,SPS08_NA49} and 
AGS \cite{AGS04} data are well described, 
the present experimental $\Xi^-/\Lambda$ ratio is underestimated 
(cf. Fig.\,\ref{xi_exc_fct}). A similar small ratio is derived \cite{Andronic09} when taking  
- instead of the calculations along the schematic freeze-out curve for 
Au+Au \cite{AndronicPBMRedlich06} - the optimum input parameters 
(i.e.\ temperature, charge and baryon chemical potentials, strangeness correlation volume) 
following from the best fit to all HADES particle yields in Ar+KCl at 1.76$A$~GeV 
\cite{PhD_Schmah,hades_kpm_phi_09}. 
Finally, recent predictions within a transport approach \cite{CheMingKo04} 
(soft EoS with incompressibility $K_0=194$\,MeV at normal nuclear matter density) 
yield a $\Xi^-/\Lambda$ ratio of a few times $10^{-4}$, 
comparable to the statistical model.

In summary, we observed the production of the doubly strange cascade hyperon $\Xi^-$ 
in collisions of Ar+KCl at 1.76$A$~GeV with a significance of about five. 
For the first time, using the HADES detector at SIS18/GSI, 
this hyperon was measured below the threshold in free nucleon-nucleon collisions, 
i.e.\ at $\sqrt{s_{NN}}-\sqrt{s_{thr}}=-640$\,MeV. 
Comparing the experimental $\Xi^-/(\Lambda+\Sigma^0)$ ratio to the predictions of a 
statistical model and a transport approach, both ones underestimate the 
experimental ratio. We conclude that I) 
the conditions for the applicability of present statistical models 
might be not fulfilled for such rare-particle production in small systems far below threshold, 
and that II) in transport approaches a better understanding is necessary of the 
strangeness-exchange reactions conjectured as the dominant process for cascade 
production below and close to threshold. 
Other explanations for the unexpectedly high $\Xi^-$ yield, like 
modifications of strange hadrons in the nuclear medium, should be investigated. 

The HADES collaboration gratefully 
acknowledges the support by BMBF grants 06TM970I, 06GI146I, 06F-140,
and 06DR135 (Germany), by GSI (TM-FR1, GI/ME3, OF/STR), by grants GA
AS CR IAA100480803 and MSMT LC 07050 (Czech Republic), by grant KBN
5P03B 140 20 (Poland), by INFN (Italy), by CNRS/IN2P3 (France), by 
grants MCYT FPA2000-2041-C02-02 and XUGA PGID T02PXIC20605PN 
(Spain), by grant UCY-10.3.11.12 (Cyp\-rus), by INTAS grant 
06-1000012-8861 and EU contract RII3-CT-2004-506078.

\end{document}